\documentclass[prl,aps,epsf,twocolumn,superscriptaddress,showpacs,10pt,]{revtex4-2}
\usepackage{graphicx,amsfonts,times,bm,amsmath,verbatim,color,array}
\usepackage{xcolor}
\usepackage{algorithmic}
\usepackage[colorlinks,urlcolor=blue,citecolor=blue,linkcolor=blue]{hyperref}

\usepackage{multirow}
\usepackage[all]{hypcap} 
\usepackage{xspace}
\usepackage{amssymb}
\usepackage{appendix}
\usepackage{pifont}
\usepackage{booktabs}
\usepackage{etoolbox} 
\usepackage{lipsum} 
\usepackage[capitalize]{cleveref}

\AtBeginDocument{%
  \heavyrulewidth=.08em
  \lightrulewidth=.05em
  \cmidrulewidth=.03em
  \belowrulesep=.65ex
  \belowbottomsep=0pt
  \aboverulesep=.4ex
  \abovetopsep=0pt
  \cmidrulesep=\doublerulesep
  \cmidrulekern=.5em
  \defaultaddspace=.5em
}

\begin{document}

\title{Gauge Field Induced Unconventional Skin Effect in Spinful Non-Hermitian Systems}

\author{Moirangthem Sanahal}
\email{moirangthem24$_$rs@phy.nits.ac.in}

\author{Subhasis Panda}
\email{subhasis@phy.nits.ac.in}

\author{Snehasish Nandy}
\email{snehasish@phy.nits.ac.in}

\affiliation{Department of Physics, National Institute of Technology Silchar, Assam 788010, India}

\begin{abstract}

The non-Hermitian skin effect (NHSE), a hallmark of non-Hermitian systems, stems from the topological nature of complex energy spectra, typically characterized by a non-zero spectral winding number. Beyond the spinless frameworks considered so far, here we realize a generic, tunable spinful NHSE in a 1D tight-binding lattice endowed with spin-dependent Abelian gauge fields. With proper tuning of the gauge parameter, we uncover an emergence of bidirected, spin-polarized zero-winding skin states, appearing in the absence of transpose-type time-reversal symmetry ($\mathcal{TRS}^{\dagger}$) and featuring scale-restricted localization with non-Bloch spectral stability. While clearly distinct from the known $\mathbb{Z}_2$ and Critical NHSEs, these unconventional skin states evolve into them upon enforcing $\mathcal{TRS}^\dagger$ and introducing inter-spin coupling via magnetic fields, respectively. The magnetic field further drives a transition from a bidirectional to a unidirectional skin configuration. Our work unifies the previously known zero-winding NHSEs within a broader framework and provides experimentally accessible routes for realization in photonic and ultracold atomic systems with synthetic gauge fields.
\end{abstract}

\maketitle

{\color{blue}{\em Introduction}}---The non-Hermitian skin effect (NHSE)~\cite{anomalousedgestates,edgestatesandtopinvariants,topologicalorigin,aperspective,areview,NH_topological_phenomena:A_review,topologicalNHSE-frontiersinPhysics,tutorial:non-hermitian_band_winding_&_NHSE,NH_topology_and_EP_geometry}, a unique feature of non-Hermnitian (NH) systems, characterized by the accumulation of bulk states at the system boundaries attributes itself to a complete breakdown of the \textit{conventional bulk-boundary correspondence}~\cite{Biorthogonal_BBC,NH_chern_bands,topPhasesofNHsystem,whydoesBBCfail,anomalousedgestates,edgestatesandtopinvariants}. Extensive theoretical efforts have been devoted to understand NHSE across a wide range of NH systems~\cite{areview,NH_topological_phenomena:A_review,topologicalNHSE-frontiersinPhysics,aperspective,tutorial:non-hermitian_band_winding_&_NHSE,NH_topology_and_EP_geometry,GKD_fingerprint,thermal_NHSE}. It is now well-established that the NHSE in one-dimension (1D) stems from an unconventional topological origin in terms of complex energies of the system, typically captured by a non-zero spectral winding number (WN)~\cite{topologicalorigin,correspondence,R_Jan-Slager}. On the contrary, recent studies have uncovered two intriguing classes of NHSE associated with \textit{zero winding number}: (i) $\mathbb{Z}_2$ skin effect --- transpose-type time-reversal symmetry $(\mathcal{TRS}^{\dagger})$ protected skin effect that is bidirectional in character and governed by a $\mathbb{Z}_2$ invariant~\cite{topologicalorigin,topPhasesofNHsystem,aperspective,internalsymmetry} and (ii) critical NHSE (cNHSE) --- arising  whenever subsystems with dissimilar non-reciprocal accumulations are coupled ~\cite{critical_NHSE_Nature,scaling_rule_for_cNHSE,Universal_competitive_spectral_scaling_cNHSE,NHSE_in_arbitrary_dimensions}. The latter, characterized by a discontinuous transition in the thermodynamic limit is accompanied by its trademark features of system-size dependent localization~\cite{scaling_rule_for_cNHSE} and non-Bloch spectral instability~\cite{Universal_competitive_spectral_scaling_cNHSE}. In line with the theoretical developments, recent years have also witnessed experimental progress of NHSE in various platforms ranging from classical systems~\cite{twisted_winding_topology,non_reciprocal_robotic_metamaterials}, electrical circuits~\cite{Reciprocal_SE_in_topoelectrical_circuit,Gen_BBC_in_topoelectrical_circuit,NHSE_in_NH_electrical_circuit}, and ultracold atoms~\cite{NHSE_in_UCatoms} to photonics~\cite{manipulate_NHSE_by_ORR,2D_NHSE_synthetic_photonic_lattice,topological_funelling,NH_BBC_in_quantum_dynamics,generating_arbitrary_topological_winding}, thereby enabling its promising applications in realizing versatile directional transports~\cite{NH_transparency_&_one-way_transport_in_low_D,Robust_light_transport-scientificreport,nonadiabrobustexc,NH_bidirectional_robust_transport} and enhanced sensing~\cite{topological_funelling,enhanced_sensitivity_via_NH_topology}.

Recent works~~\cite{Control_NHSE_via_SGF,synthetic_non_abelian,nonAbelian_self-healing} have demonstrated the NHSE in 1D systems endowed with synthetic non-Abelian gauge fields; effectively elevating its internal degrees of freedom to realize and manipulate bidirectional NHSE. These advances have opened new directions and generated considerable interests within the community. However, the study of NHSE in spinful systems subject to Abelian gauge fields remains largely unexplored. Spin degrees of freedom, besides being intrinsic to realistic systems, are also externally controllable via magnetic fields. Their fundamental role in relevant aspects, including topological insulators and quantum spin Hall effect~\cite{quantum_spin_hall,asbothTopInsulators}, underscores the broader relevance of spinful systems. The choice of Abelian gauge fields further enables independent control and manipulation of individual spin sectors, rendering it particularly suitable for spintronic applications~\cite{spintronics:review}. In this spirit, the investigation of the NHSE in spinful NH systems represents a compelling frontier, with the potential to unveil novel physical phenomena and significantly broaden the landscape of NH topological physics.
 
\begin{figure}[b]
    \centering
    \includegraphics[width=0.9\linewidth]{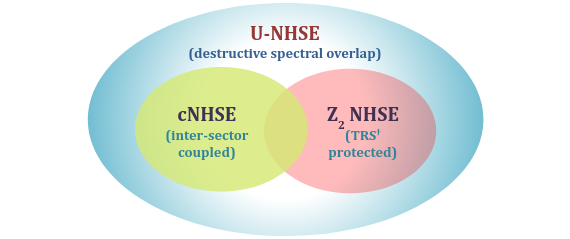}
    \caption{\small Venn diagram illustrating the broader framework of U-NHSE; encompassing both cNHSE and the $\mathbb{Z}_2$ NHSE as limiting cases under corresponding conditions. The interesection: cNHSE $\cap$ $\mathbb{Z}_2$ NHSE highlights the emergence of critical NHSE in $\mathcal{TRS}^\dagger$ protected systems as shown in~\cite{Supplementary}.}
    \label{fig:Figure1} 
\end{figure}

In this work, we study the NHSE in a 1D tight-binding lattice subjected to spin-dependent Abelian gauge fields, invoking its spin degrees of freedom. 
Interestingly, this NHSE exhibits a coexistence of two distinct types of skin modes with tunable characteristics: (i) a conventional one with non-zero WN, consistent with the well-established `winding number-skin state correspondence'~\cite{correspondence} and (ii) an unconventional one arising despite a vanishing WN and in the absence of $\mathcal{TRS}^{\dagger}$. Remarkably, the latter evades classification by both WN and the $\mathbb{Z}_2$ invariant, while also remaining fundamentally distinct from the cNHSE as it exhibits scale-restricted localization with non-Bloch spectral stability. We dub the skin effect originating from these unusual skin modes as ``unconventional NHSE (U-NHSE)''. Introducing an external magnetic field however, suppresses these unconventional skin states, giving way to cNHSE. Meanwhile, the conventional ones stay robust but undergo transition from bidirectional to unidirectional accumulation with increasing field strength. Finally, we demonstrate that the cNHSE and the $\mathbb{Z}_2$ skin effect naturally emerge as limiting manifestations of the broader framework of U-NHSE, in the presence of inter-sector coupling and $\mathcal{TRS}^\dagger$ respectively, as illustrated in Fig.~\ref{fig:Figure1}. A practical scheme for realizing U-NHSE in multi-band systems, is proposed in Fig.~\ref{fig:Figure2}, enabling the exploration of our predicted signatures in photonic platforms utilizing synthetic gauge fields.

\maketitle

\begin{figure}[b]
    \centering
    \includegraphics[width=0.46\textwidth]{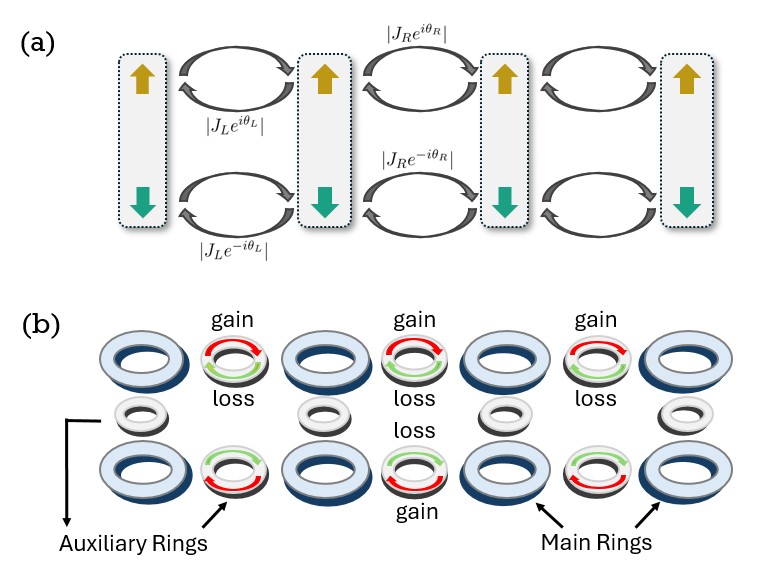}
    \caption{\small Schematic diagram of Abelian spinful 1D chain. (a) Spin-dependent coupling strengths under Abelian gauge combinations: $\mu,\nu = z$. The model replicates combination of two 1D TB chains without inter-chain coupling. (b) Scheme of the structure for realizing spinful 1D chain using optical ring resonator particularly for the case: $\text{Im}(\theta_L)>0$, $\theta_R\in\mathbb{R}$ (and $\textbf{B} \perp \hat{z}$ in case if magnetic field is applied.)}
    \label{fig:Figure2} 
\end{figure}

{\color{blue}{\em 1D spinful model}}---The generic Hamiltonian of a 1D tight-binding (TB) lattice encompassed with spin-dependent gauge fields, describing the dynamics of a spin-$\frac{1}{2}$ particle along the chain can be written as~\cite{nonabelian_lattice_gaugefields_in_phtonic...}
\begin{equation}
    \mathcal{H} = \sum_{m} (J_Le^{iA_{\mu}} c_{m}^\dagger c_{m+1} + J_R e^{iA_{\nu}} c_{m+1}^\dagger c_{m}), \label{eqn:zz Abelian real space Hamiltonian}
\end{equation}
where $J_{L (R)}$ denotes the hermitian hopping amplitude towards the left (right) nearest-neighboring site. Here, $A_{{\mu}(\nu)}$ is an arbitrary matrix-valued gauge potential of the form $\theta_{L(R)}\sigma_{\mu(\nu)}$ with $\theta_{L(R)}$ being the associated gauge flux and $\sigma_{\mu}$ $(\mu\in x,y,z)$ representing Pauli matrices acting on the spin degrees of freedom~\cite{synthesis&observation_of_nonAbelian_gauge_fields_in_real_space,light_induced_Gauge_in_ultracold_atom_(REVIEW),colloquim:artificial_gauge_fields,synthetic_non_abelian}. The gauge types $\mu,\nu$ determine the commutativity of the corresponding gauge fields, defining the system as either Abelian (commutative) or non-Abelian (noncommutative)~\cite{nonabelian_lattice_gaugefields_in_phtonic..., synthesis&observation_of_nonAbelian_gauge_fields_in_real_space,synthetic_non_abelian}. In this work, we restrict ourselves to the Abelian case by considering identical gauge types ($\mu,\nu=z$). Fig.~\ref{fig:Figure2}(a) shows a schematic diagram of the hopping processes under this gauge combination. For simplicity, we set $J_L ,J_R\ =1$ throughout, isolating the role of the Abelian gauge fields on non-Hermiticity and spin-dynamics in the system.

In the context of internal symmetries, the above system respects conjugate-type time-reversal symmetry ($\mathcal{TRS}$): $U_{T_+}\mathcal{H}^*(k) U_{T_+}^{-1} = \mathcal{H}(-k)$ with $U_{T_+}U_{T_+}^{*}=-1$ where $U_{T_+}=i\hat{\sigma}_y$ but breaks particle-hole, chiral and transpose-type time-reversal symmetries ($\mathcal{TRS}^{\dagger}$) thereby belonging to AII symmetry class~\cite{symmetry&topology,NH_topology_ib_H_matter}.

\begin{figure}[t]
    \centering
    \includegraphics[width=0.45\textwidth]{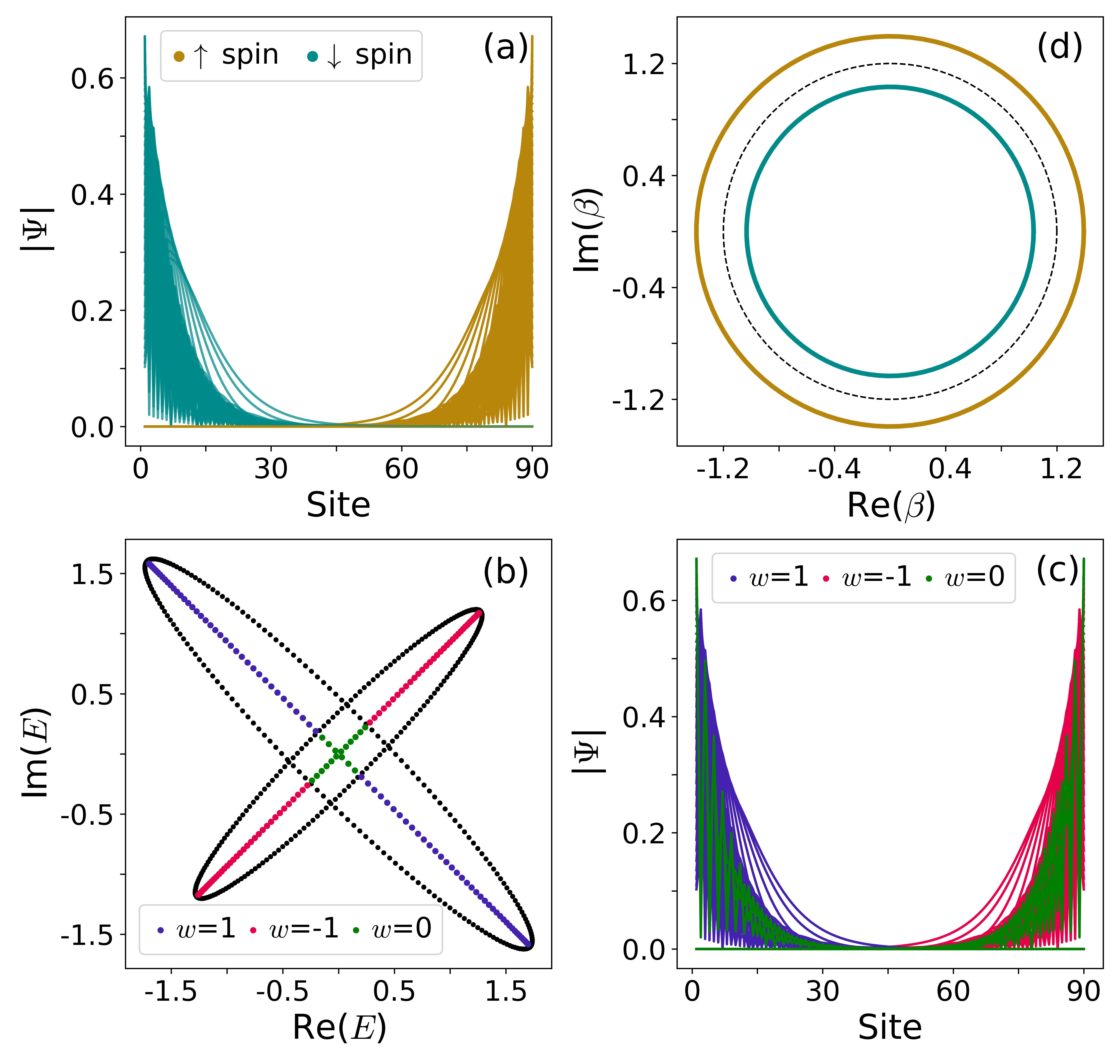}

    \caption{\small Numerical analysis of spinful NHSE. (a) Spin-polarized skin modes, with up (down) spin localizing right (left)-wards. (b) Eigenspectra: the two closed PBC (black) loops exhibit opposite windings ($w$=$\pm1$). The OBC eigenenergies enclosed within are color coded with respect to $w$. (c) Corresponding OBC eigenstates showing localization. Energies within the overlapping region (green) have zero $w$ but their corresponding eigenstates show boundary localization. (d) Non-Bloch analysis: the GBZs comprising two concentric circles explains the simultaneous existence of left-($\text{radius}<1$) and right-localized ($\text{radius}>1$) eigenstates. Here, $\theta_L=0.3i$, $\theta_R=1.5$ and $N=90$.}
    \label{fig:Figure3}
\end{figure}

\begin{table*}[t]
\renewcommand{\arraystretch}{0.75}
\setlength{\tabcolsep}{16pt}
\begin{tabular}{@{}cccccc@{}}\toprule

$\theta_L,\theta_R$ & $J_L, J_R$ & 
Condition&
NHSE nature &
Spin skin modes &
Separation \\ 

\midrule

\multirow{ 2}{*}{Imag, Real} & \multirow[c]{6}{*}[-0.25cm]{$|J_L| = |J_R|$} & ${\rm Im}(\theta_L)>0$ & \multirow[c]{6}{*}[-0.26cm]{Bidirectional} & $\uparrow$: Right, $\downarrow$: Left & \multirow[c]{6}{*}[-0.26cm]{Symmetric} \\

 & & ${\rm Im}(\theta_L)<0$ & & $\uparrow$: Left, $\downarrow$: Right & \\ \cmidrule(r){1-1}\cmidrule(lr){3-3}\cmidrule(lr){5-5} 

\multirow{ 2}{*}{Real, Imag} &  & ${\rm Im}(\theta_R)>0$ &  & $\uparrow$: Left, $\downarrow$: Right &  \\

 & & ${\rm Im}(\theta_R)<0$ & & $\uparrow$: Right, $\downarrow$: Left & \\ \cmidrule(r){1-1}\cmidrule(lr){3-3}\cmidrule(lr){5-5} 
 
\multirow{ 2}{*}{Imag, Imag} & & ${\rm Im}(\theta_L)>{\rm Im}(\theta_R)$ & & $\uparrow$: Right, $\downarrow$: Left &  \\

 & & ${\rm Im}(\theta_L)<{\rm Im}(\theta_R)$ & & $\uparrow$: Left, $\downarrow$: Right &  \\ \midrule

Real, Real &  $|J_L| = |J_R|$ & Arbitrary & No skin effect & --- & --- \\

\bottomrule
\end{tabular}
\label{tab:Table1}
\caption{\small Summary of skin effect and spin skin mode behavior under various parameter regimes for the Abelian case: $\mu,\nu=z$.}
\end{table*}

{\color{blue}{\em Conventional bidirectional skin effect}}---In our system, the non-Hermiticity induced solely by Abelian gauge fields gives rise to a bidirectional NHSE even in the presence of reciprocal hopping. This markedly differs from the behavior observed with non-Abelian gauge fields~\cite{synthetic_non_abelian}. However, it requires at least one of the gauge fields to be \textit{imaginary}~\cite{Supplementary}. Starting with a uniform imaginary $\sigma_z$ gauge field ($\theta_L \in \mathbb{C}$), the system shows a \textit{spin-polarized bidirectional symmetric NHSE} as depicted in Fig.~\ref{fig:Figure3}(a) with spin-up (down) modes accumulating at the right (left) end under open boundary conditions (OBC). This contrasts with the previous study where spin-up states propagate left and spin-down states right~\cite{spinful}. The nature of NHSE in the presence of different choices of Abelian gauge fields for a spinful 1D chain is shown in Table. \ref{tab:Table1} (also refer~\cite{Supplementary}).   

The imaginary gauge field effectively modifies the spin-dependent hopping amplitudes unlike real gauge fields~\cite{Robust_light_transport-scientificreport,NH_transparency_&_one-way_transport_in_low_D,nonadiabrobustexc}. It leads to independently dissimilar non-reciprocity in the two decoupled spin subspaces, physically explaining the emergence of spin-polarized bidirectional NHSE. For instance, with imaginary $\theta_L$ (i.e. $i\rm Im(\theta_L))$ and real $\theta_R$, the leftward, rightward hopping amplitudes are modified as $|J_L e^{i\theta_L}| \to |J_L e^{- \rm Im(\theta_L)}|$, $|J_R e^{i\theta_{R}}|\to |J_R|$ in the up-spin sector and $|J_L e^{-i\theta_L}| \to |J_L e^{\rm Im(\theta_L)}|$, $|J_R e^{-i\theta_{R}}| \to |J_R|$ in the down-spin sector. 

To delve into further, we rewrite the real and imaginary part of the Bloch eigenvalues as $[ \text{Re}(E_{\pm}(k)), \text{Im}(E_{\pm}(k))]^{T} = \mathcal{R_{\pm}} [\text{cos}k, \text{sin}k]^T$ with $\mathcal{R}_\pm$ as a transformation matrix and $\pm$ representing up (down)-spin sector~\cite{Supplementary}. The Bloch eigenspectrum, constituted by $( \text{Re}(E_{\pm}(k)), \text{Im}(E_{\pm}(k)))$ as its loci on the complex plane for $k\in[0,2\pi)$ traces two origin-centered ellipse tilted along $43^{\circ},-43^{\circ}$ lines. The ellipse corresponding to up (down)-spin sector spans itself anticlockwise (clockwise) as parameter $k$ sweeps across $0$ to $2\pi$ leading to opposite senses of winding. Meanwhile, the above Hamiltonian under OBC has the block-diagonal form $H_{\uparrow}\otimes H_{\downarrow}$ with each block resembling Toeplitz tridiagonal form~\cite{toeplitz,nonBlochBandTheoryandBBC},  allowing an analytic description of OBC eigenvalues for each spin sector:  $E_{n,\pm}^{OBC}= e^{\pm i(\theta_L+\theta_R)/2}\ (2 \sqrt{J_LJ_R} \cos(\frac{n\pi}{N+1}))$. Here, $n$ is the eigenvalue index and $N$ denotes the physical system size. The expression maps two straight, intersecting line segments; enclosed independently within the two distinct $E_{\pm}(k)$ loops~\cite{Supplementary}. This explains the spin polarization in the bidirected skin modes (see Fig.~\ref{fig:Figure3}(a)).

We further numerically compute both the eigenspectra and WN. Under periodic boundary conditions (PBC), the eigenspectra as shown in Fig.~\ref{fig:Figure3}(b) consists of two closed loops, exhibiting point-gap, whereas under OBC, it shows two open arcs. These features points towards an inevitable occurrence of NHSE ~\cite{topologicalorigin}. The two closed loops independently do carry opposite WN, $w=\pm1$, representing anti-clockwise (clockwise) winding~\cite{topologicalorigin,topPhasesofNHsystem,correspondence}. The opposite $w$ explains the simultaneous existence of left- $(w=1)$ and right-localized $(w=-1)$ eigenstates in the open system as depicted in Fig. \ref{fig:Figure3}(c). 

{\color{blue}{\em Spin-polarized unconventional skin effect}}---In addition to $w=\pm1$, we find states with $w=0$ arising from a destructive overlap of the two closed loops (Fig.~\ref{fig:Figure3}(b)) --- a feature exclusive to multi-band systems. Remarkably, these eigenstates exhibit (bi-)localization (Fig.~\ref{fig:Figure3}(c)) revealing a contextual breakdown of both the established (i) `winding number-skin state correspondence' in multi-band scenario~\cite{correspondence,topologicalorigin,NHSE_in_arbitrary_dimensions} and (ii) $\mathbb{Z}_2$ invariant correspondence as the system lacks $\mathcal{TRS}^\dagger$~\cite{topologicalorigin}. We term this behavior the ``\textit{unconventional NHSE} (U-NHSE)'', broadly encompassing all zero-winding NHSEs arising from destructive spectral overlap. Even more interestingly, these bilocalized $w=0$ states are spin-polarized - up (down) spin localizing at the right (left) boundary as shown in Fig. \ref{fig:Figure3}(a). Their directionality adheres to the same convention (Toeplitz tridiagonal form analysis) as described earlier, reflecting a subtle retention of individual-band topology underneath, with the WN at the band-resolved level capturing essential aspects of localization. This retention serves as the origin for the inherent bidirectionality of U-NHSE, which unlike $\mathbb{Z}_2$ skin effect need not be symmetric. Thus, our findings do not imply a breakdown of the correspondence itself, but rather its contextual limitations in multi-band systems. Furthermore, it is worth emphasizing that the $\mathbb{Z}_2$ skin effect emerges as a special case of the U-NHSE upon enforcing $\mathcal{TRS^{\dagger}}$~\cite{Supplementary}, underscoring the broader generality of U-NHSE.
 
Due to the block diagonal form of the Hamiltonian, the GBZ trajectories are the union of the individual sub-GBZs corresponding to each diagonal entry \cite{auxillaryGBZ,tutorial:non-hermitian_band_winding_&_NHSE}. Fig. ~\ref{fig:Figure2}(d) depicts the GBZ trajectories, which are two concentric circles of radii $e^{-{\rm Im}(\theta_L)/2}$ and $e^{{\rm Im}(\theta_L)/2}$ respectively. The trajectory inside (outside) the conventional BZ indicates the left (right) localized eigenstates with $w=1~(w=-1)$. 

{\color{blue}{\em Tunable characteristics}}---The obtained U-NHSE can be effectively tuned via the imaginary gauge field parameter since it modifies the hopping amplitude. As pointed out earlier, for $\rm Im(\theta_L)>0$ (source) $(<0$ (sink)), the up-spin states localize at the right (left) end, while spin-down states localize at the left (right) edge. This enables directional control over the spin skin modes. Moreover, the localization length can as well be controlled via $e^{\rm Im(\theta_L)}$. When both the gauge fields are imaginary, the direction and strength of localization depend on the relative difference $\rm Im(\theta_L)-\rm Im(\theta_R)$ and exponential factor $e^{\rm Im(\theta_L)-\rm Im(\theta_R)}$ respectively~(Table.~\ref{tab:Table1}) \cite{Supplementary}. Interestingly, this configuration also allows for a complete suppression of the skin effect by tuning the imaginary components to be equal i.e., $\rm Im(\theta_L) = \rm Im(\theta_R)$; a case absent in other configurations. It is important to note that these tunable characteristics hold equally true for conventional $(w = \pm1)$ spin skin modes. Further asymmetry in localization can be introduced via non-reciprocal hopping $(J_L\neq J_R)$, which, if added on top of the gauge-induced NHSE, can drive a crossover from bidirectional to unidirectional accumulation~\cite{Supplementary}.

{\color{blue}{\em Critical NHSE induced by Zeeman field}}--- Next, we explore the effect of an external magnetic field (\textbf{B}) to the system. Given 1D nature, the unavailability of minimal room for cyclotron motion and consequent orbital effects allows only Zeeman contribution to arise from direct coupling between magnetic field and the spin. 

\begin{figure}[t]
    \centering
    \includegraphics[width=0.45\textwidth]{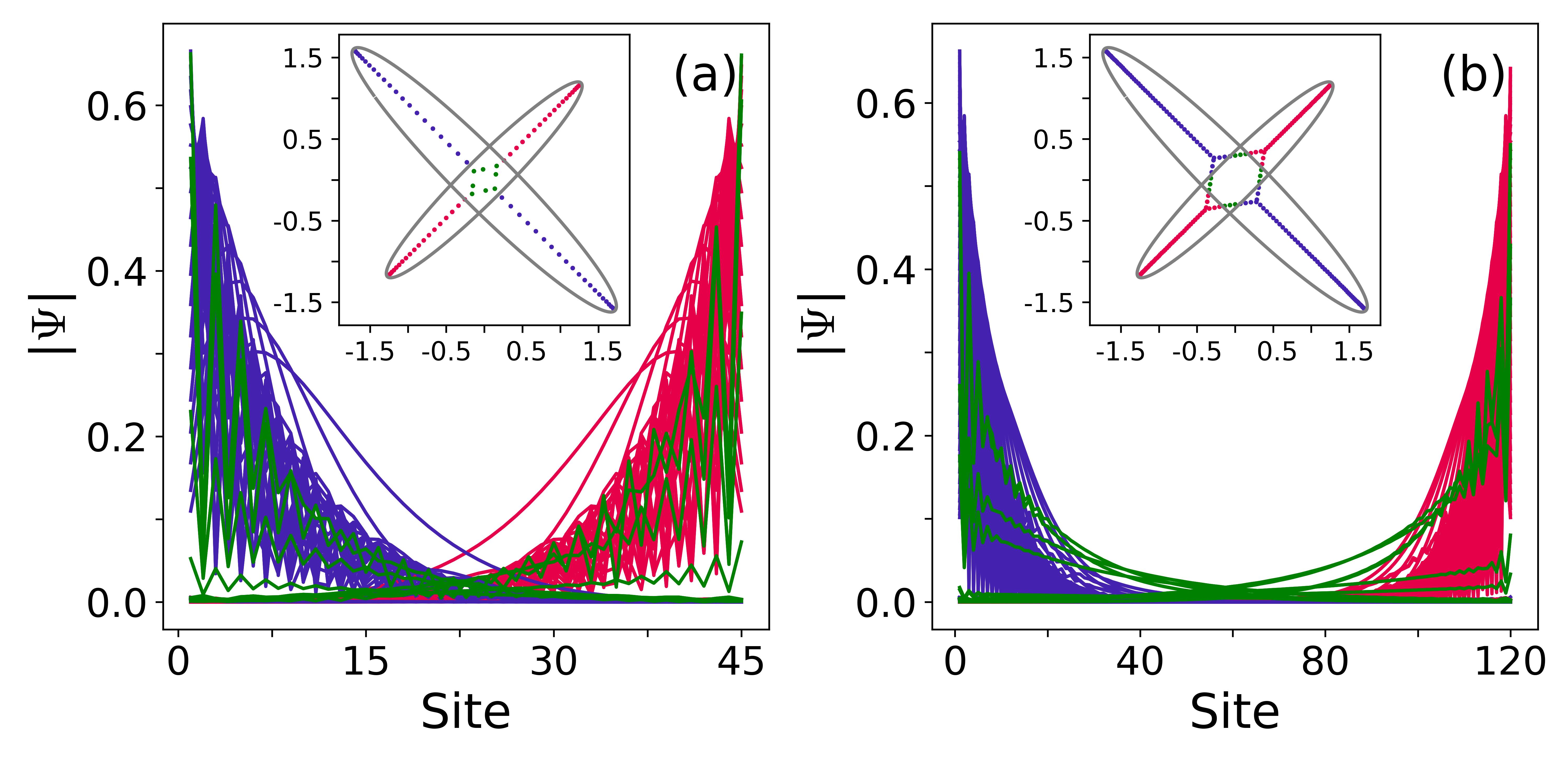}
    \caption{\small Critical NHSE showing system-size dependence. Eigenstates and corresponding eigenspectrum (inset) with system sizes (a) $N=45$ and (b) $N=120$ under external magnetic field $B_x=0.01$. The gray closed loops represent the PBC spectrum. The OBC energies (green) within the overlapping region have $w=0$ and the red (violet) ones outside have $w=\mp1$. Other parameters are same as in Fig.~\ref{fig:Figure3}.}
    \label{fig:Figure4}
\end{figure}

The Zeeman term ($\textbf{B}.\bm{\sigma}$) affects the spin sectors only if it has an in-plane component for otherwise $[H,\sigma_z]=0$; preserving the spin-polarized states. The in-plane magnetic field (for instance, $B_x \hat{i})$ physically induces onsite inter-spin coupling in our system. This coupling between two spin sectors having dissimilar non reciprocities leads to the emergence of cNHSE. In particular, the unconventional ($w=0$) states now show size-dependent localization where the localization length varies with system size. Concurrently, the part of the non-Bloch spectrum within the overlapping region ($w=0$) protrudes outwards as system size increases. In contrast, the spectral regions with $w\neq0$ remain largely unaffected. Fig. \ref{fig:Figure4} displays the eigenspectrum and corresponding eigenstates for different system sizes under finite coupling. The scale free localization and non-Bloch spectral instability of the $w=0$ states with respect to system size are consistent with the hallmark features of cNHSE and tied to the zero  winding~\cite{critical_NHSE_Nature,NHSE_in_arbitrary_dimensions}. Thus, while U-NHSE generically defines the existence of zero-winding skin states via destructive spectral overlap, the emergence of cNHSE upon activating inter-sector coupling builds atop this framework, once again highlighting the foundational generality of U-NHSE.

\begin{figure}[t]
    \centering
   
    \includegraphics[width=0.47\textwidth]{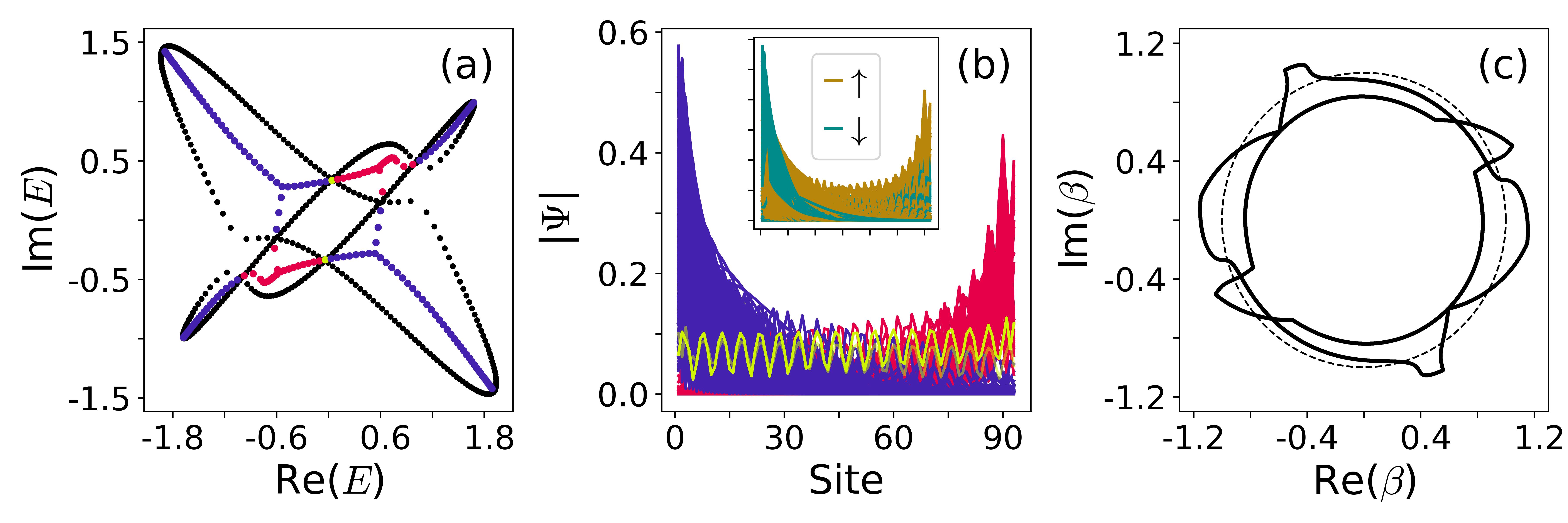}
    \quad
    \includegraphics[width=0.47\textwidth]{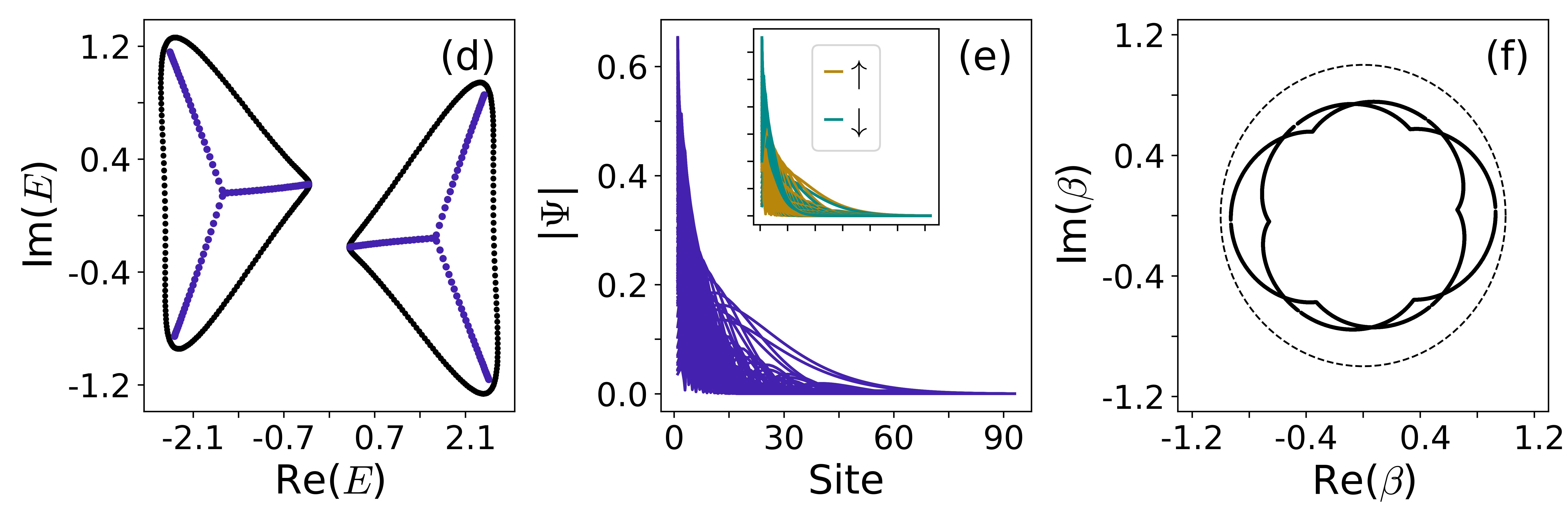}
    \caption{\small Evolution of spin-polarized bidirectional NHSE to unidirectional NHSE under $B_x$. Eigenspecrtum, corresponding OBC eigenstates and GBZ at (a-c) $B_x=0.85$ and (d-f) $B_x=1.7$. The main figures in (b), (e) are color coded based on $w$, whereas the corresponding insets show the same data, but with a different color scheme to distinguish the spin components. (c) The intersection of GBZ with BZ (dotted unit circle) indicates the existence of extended, $w=0$ states in (b). (f) The enclosure of the entire GBZ within BZ explains the existence of only left localized, $w=1$ skin states as in (c). Parameters include N=93 and others, same as in Fig.~\ref{fig:Figure3}.}
    \label{fig:Figure5}
\end{figure}

{\color{blue}{\em Bidirectional to Unidirectional Skin Effect}}---With increasing $B_x$ (or $B_y$), the system undergoes a transition from bidirectional to unidirectional NHSE as illustrated in Fig. \ref{fig:Figure5}. In the low field regime, the PBC eigenspectrum remains almost the same whereas the number of critical skin modes decreases. As the field strength increases, the two PBC loops distort, enclosing more number of OBC states with $w=1$ (Fig.~\ref{fig:Figure5}(a)) and eventually separate into two distinct loops with identical WN (Fig.~\ref{fig:Figure5}(d)). The subsequent enclosure of entire OBC spectrum within both PBC loops of identical winding explains the transition of bidirectional NHSE into unidirectional NHSE~\cite{spinful}. The directionality of the localization depends on choice of gauge configuration. For instance, when $\theta_L\in \mathbb{C},\ \theta_R\in \mathbb{R}$, states localize leftwards; as dictated by the relation: $\text{cosh}(\text{Im}(\theta_L))>\text{cos}(\theta_R)$ (see supplementary~\cite{Supplementary} for detail). It is to be noted that the OBC eigenenergies in Fig. \ref{fig:Figure5}(a) corresponding to $w=0$ exhibit extended states - an outcome attributed to twisted winding topology~\cite{twisted_winding_topology}, unlike the case in Fig. \ref{fig:Figure3}(c) where the $w=0$ states show localization due to spectral overlap. It is also observed that the application of an external magnetic field can induce tunable conventional skin states even in the regime where both gauge fields are real $\theta_L,\theta_R \in \mathbb{R}$~\cite{spinful}. 

The GBZs under different magnetic field strengths, $B_x$ are shown in Figs.~\ref{fig:Figure5}(c,f). Since $B_x$ introduces spin-flip hopping, the Bloch Hamiltonian, no longer in block diagonal form deforms the GBZ from concentric circles. It is clear from the figure that in the weak field regime, the deformations exhibit certain intersections with the BZ, delineating the presence of extended states. With increasing strength, as the states transit to a unidirected localization, the corresponding GBZs subsequently intrude into the BZ, signaling the unidirectional accumulation with absence of extended states. In view of this, the intersection between GBZ and BZ could be a diagnostic feature for distinguishing $w=0$ extended states from $w=0$ unconventional skin states (Fig.~\ref{fig:Figure3}(b)).

We extend our analysis to Abelian cases with $\mu=\nu \neq z$ and find that key features of different spinful NHSEs persist, except spin unpolarizability~\cite{Supplementary}. This arises from spin-flip hoppings via off-diagonal elements in the associated Pauli matrices. Moreover, an external magnetic field orthogonal to $\mu$ induces effects analogous to case for $B_x$ (or $B_y$) in $\sigma_z$-type Abelian gauge fields.

{\color{blue}{\em Discussions.}}---In summary, we study the spinful NHSE using a 1D TB chain threaded with spin-dependent Abelian gauge fields and uncover a generic class of bidirected NHSE associated with zero WN. Originating from a destructive spectral overlap, this unconventional NHSE not only manifests beyond the conventional frameworks of both WN and the $\mathbb{Z}_2$ invariant, but also exposes their limitations through band-resolved topology that accounts for the spin polarization of observed skin states. While the former becomes inadequate in multi-band context, the latter remains strictly constrained to $\mathcal{TRS}^\dagger$, highlighting the need to transcend conventional WN or $\mathbb{Z}_2$-invariant correspondence in multi-band settings. We further show that introducing an external magnetic field and enforcing $\mathcal{TRS}^\dagger$ evolves the U-NHSE into cNHSE and $\mathbb{Z}_2$ NHSE, marked by their hallmark features, respectively. The U-NHSE thus provides a broader framework encompassing both the $\mathbb{Z}_2$ and critical NHSEs as limiting cases. These behaviors offer a tunable platform for spin-selective localization and highlight the role of magnetic field in control and manipulation of them in spinful NH systems.

Our findings can be realized in diverse scenarios, viz. photonic systems using optical ring resonators with controlled gain and loss mediums as presented in~Fig. \ref{fig:Figure2}(b)~\cite{manipulate_NHSE_by_ORR,2D_NHSE_synthetic_photonic_lattice,Robust_light_transport-scientificreport,NH_transparency_&_one-way_transport_in_low_D}. Here, the gain (red) and loss (green) processes along the auxiliary rings mimic the anisotropic coupling between main rings (lattice sites), while the uniform auxiliary rings linking the two chains replicate the magnetic field effect. It should be stimulating
to extend our findings to higher dimensions, where NHSE becomes a rather universal property of non-Hermitian systems~\cite{universal}.

\textcolor{blue}{\textit{Note.}}---During the course of this work, we noticed a preprint~\cite{spinful} addressing NHSE in spinful systems with partial overlap. While both studies consider similar 1D TB chain and gauge configurations, our analysis distinctly elucidates the zero-winding spectral overlap and its emergent consequences.
 
\textcolor{blue}{Acknowledgments}---We thank Kohei Kawabata for his valuable insights and suggestions. We also thank Heming Wang, Shanhui Fan and Adithya JD for helpful discussions. We acknowledge the computing resources of `PARAM SHAVAK' at Computational Condensed Matter Physics Lab, Department of Physics, NIT Silchar.


\bibliographystyle{unsrt}
\bibliography{ref}

\newpage

\end{document}